\documentclass{aastex631}
\usepackage{textcomp}
\usepackage{graphicx}
\usepackage{amssymb}
\usepackage{amsmath}
\usepackage{epstopdf}
\usepackage{gensymb}
\usepackage{natbib}
\usepackage{appendix}
\usepackage{tabu}
\usepackage{color}
\usepackage{multirow}
\usepackage{booktabs}
\usepackage{ulem}
\usepackage{threeparttable}
\definecolor{darkblue}{rgb}{0.17, 0.49, 0.72}
\definecolor{darkgreen}{rgb}{0.0, 0.4, 0.0}
\newcommand{\sref}[1]{Section~\ref{#1}}
\newcommand{\fref}[1]{Figure~\ref{#1}}
\newcommand{\eqn}[1]{Equation~(\ref{#1})}
\newcommand{\tab}[1]{Table~\ref{#1}}

\newcommand{\sind}{\mbox{$S$-index}}
\newcommand{\ca}{\mbox{Ca\,{\sc ii} H and K}}

\defcitealias{Sowmyaetal2021}{Paper I}

\begin{document}
\title{Modeling Stellar \ca{} Emission Variations: Spot Contribution to the \texorpdfstring{$S$}{}-index.}
\received{September 13, 2023}
\submitjournal{ApJL}
\shortauthors{Sowmya et al.}

\correspondingauthor{K.~Sowmya}
\email{krishnamurthy@mps.mpg.de}

\author[0000-0002-3243-1230]{K.~Sowmya}
\affiliation{Max-Planck-Institut f\"ur Sonnensystemforschung, Justus-von-Liebig-Weg 3, 37077 G\"ottingen, Germany}

\author[0000-0002-8842-5403]{A.~I.~Shapiro}
\affiliation{Max-Planck-Institut f\"ur Sonnensystemforschung, Justus-von-Liebig-Weg 3, 37077 G\"ottingen, Germany}

\author[0000-0003-2088-028X]{L.~H.~M. Rouppe van der Voort}
\affiliation{Institute of Theoretical Astrophysics, University of Oslo, PO Box 1029, Blindern 0315 Oslo, Norway}
\affiliation{Rosseland Centre for Solar Physics, University of Oslo, PO Box 1029, Blindern 0315 Oslo, Norway}

\author[0000-0002-1377-3067]{N.~A.~Krivova}
\affiliation{Max-Planck-Institut f\"ur Sonnensystemforschung, Justus-von-Liebig-Weg 3, 37077 G\"ottingen, Germany}

\author[0000-0002-3418-8449]{S.~K.~Solanki}
\affiliation{Max-Planck-Institut f\"ur Sonnensystemforschung, Justus-von-Liebig-Weg 3, 37077 G\"ottingen, Germany}

\begin{abstract}
The \sind{} is a measure of emission in the \ca{} lines and is a widely used proxy of stellar magnetic activity. It has been assumed until now that the \sind{} is mainly affected by bright plage regions in the chromosphere. In particular, the effect of starspots on the \sind{} has been neglected. In this study we revisit this assumption. For this we analyze high-resolution observations of sunspots recorded in the Ca\,{\sc ii} H spectral line at the Swedish 1-m Solar Telescope and determine the contrast of spots with respect to the quiet surroundings. We find that the Ca\,{\sc ii} H  line core averaged over whole sunspots (including superpenumbrae) is brighter than in the quiet surroundings and that the spot contrast in the line core is comparable to the facular contrast. This allows us to get a first estimate of the influence of spots on the \sind{}. We show that spots increase the \sind{}. While this increase is quite small for the Sun, it becomes significantly larger for more active stars. Further, we show that the inclusion of the contribution of spots to the \sind{} strongly affects the relationship between the \sind{} and stellar disk area coverages by spots and faculae, and present the new relations.
\end{abstract}


\section{Introduction}
\label{sec:intro}
Concentrated magnetic fields on the photospheres of late-type stars form dark spots and bright faculae. Such magnetic fields extend to the chromosphere where they cause additional non-thermal heating and lead to the formation of chromospheric magnetic features. One of the most exciting manifestations of appearance and disappearance of magnetic features on the visible stellar surface (either due to their evolution or due to the stellar rotation) is variations of stellar brightness. Namely, the photospheric features cause variations observed in broadband photometric passbands, while the chromospheric features are responsible for variability in strong spectral lines like \ca{} which are amongst the most widely used proxy for stellar magnetic activity \citep[see, e.g.,][and references therein]{HALL_LRSP}.

Brightness variations are very well studied for the Sun \citep{Sami2013,Ermolli2013}. In the visible and infrared spectral domains they are affected by the interplay of photospheric spots and faculae \citep{Sami2013, Sashaetal2016}. At the same time the variability of the solar \ca{} emission is mainly driven by plages, which are the chromospheric counterpart of faculae (so that the same concentration of magnetic fields causes faculae in the photosphere and plages in the chromosphere). The spot contribution to solar \ca{} variability is believed to be small  \citep[][hereafter Paper I]{Sashaetal2014,Sowmyaetal2021} because plages are very bright in \ca{} lines \citep[e.g. see Figure 1 of][]{Skumanichetal1984} and on the Sun their area coverages are significantly larger than that of spots. As a result solar \ca{} emission is even often considered as a proxy of facular contribution to solar irradiance variability \citep{Lean2000, Walton2003, Chapman2013, Berrilli2020}. However, the role of spots in \ca{} variability of other stars has not been investigated before.

Stellar \ca{} emission measurements date back to Mount Wilson Observatory's HK project in the 1960s which led to the establishment of the well known chromospheric activity indicator called the \sind{} \citep{Vaughanetal1978}, which is proportional to the summed fluxes in the H and K passbands covering the cores of the \ca{} lines, normalized to the summed fluxes in the nearby pseudo-continuum R and V passbands \citepalias[see][for the definition of the passbands and \sind{}]{Sowmyaetal2021}. The HK project was followed by Lowell observatory's monitoring of the chromospheric activity simultaneously with photometric brightness \citep{Lockwoodetal1992}. A combination of the data from this survey and the HK project revealed that old and less active stars like the Sun showed a direct correlation between the photometric brightness and the chromospheric activity while young active stars showed an anti-correlation \citep{Lockwoodetal1992,Lockwoodetal2007,Radicketal1998,Radicketal2018}, indicating a transition from faculae-dominated to spot-dominated brightness variability. This observed transition was explained to be due to surface coverages of spots growing much faster with activity than that of faculae \citep{Foukal1993,Chapmanetal1997,Foukal1998,Sashaetal2014,Nemecetal2022}. Thus, for more active stars, spots are expected to have a stronger contribution to the \sind{} variability, prompting  a quantitative assessment of their effect. 

The main hurdle in assessing the impact of spots on stellar \sind{} values is the absence of any reliable information on spot chromospheres. Since direct observations of starspots in \ca{} lines are not yet feasible the most straightforward way to circumvent this hurdle is to utilize solar observations. Such observations have very recently become available thanks to the CHROMospheric Imaging Spectrometer (CHROMIS) on the Swedish 1-m Solar Telescope \citep[SST;][]{Scharmeretal2003} which observes the Sun with unprecedented spatial and good spectral resolution. Here we analyze the SST data of three sunspots in the Ca\,{\sc ii} H line at 3968.47\,\AA{} to determine the brightness of chromospheric counterparts of spots. The details of the data analysis and modelling approach are given in \sref{sec:methods}. Making use of the observed sunspot contrast, we evaluate the spot contribution to the \sind{} values of the Sun and active Sun-like stars. We present our findings in \sref{sec:res} and conclusions in \sref{sec:conclu}.

\begin{figure*}[ht!]
    \centering
    \includegraphics[scale=0.52]{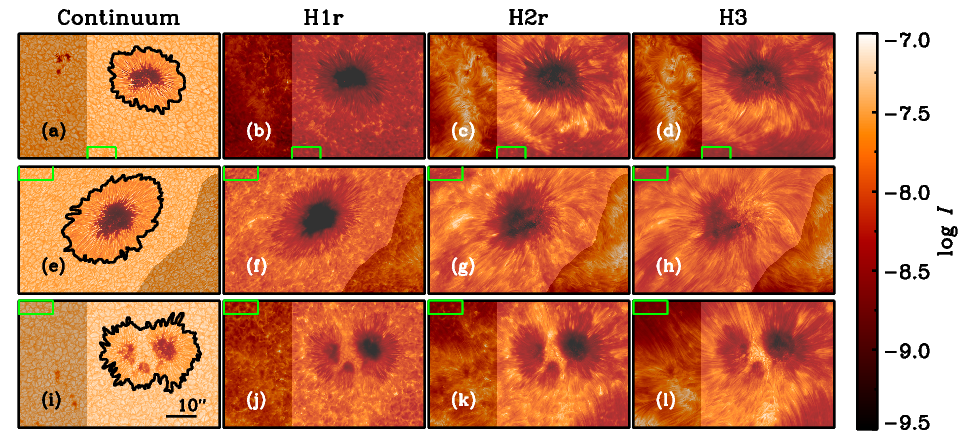}
    \caption{Intensity images of the three sunspots scanned on 07-08-2020 at $\mu=0.83$ (top row), on 22-06-2021 at $\mu=0.63$ (middle row), and on 07-07-2022 at $\mu=0.96$ (bottom row). The images cover an angular area of roughly 67''$\times$42''. The sunspots are shown in the continuum at $\sim4001$\,\AA{} (first column), and at wavelengths corresponding to H1r (second column), H2r (third column) and H3 (fourth column) as marked in \fref{fig:profcombined}. The regions shown with lower transparency mask the areas contaminated by bright facular and plage patches. These regions are excluded from the analysis. Green boxes in each panel indicate the patches used for calculating the quiet Sun intensity profiles shown in \fref{fig:profcombined}. Black contours in the first column mark the boundary of the sunspot. See text for details.}
    \label{fig:intcombined}
\end{figure*}

\section{Methods}
\label{sec:methods}
\subsection{Modeling stellar chromospheric activity}
\label{ssec:model}
In \citetalias{Sowmyaetal2021}, we modeled the \sind{} for stars with solar activity level by neglecting spots since their surface coverages are much smaller than those of faculae. Spots were assumed to affect the fluxes in the H and K as well as R and V passbands in the same way so that their effect would vanish when the flux ratio is taken. Here, the goal is to determine the \sind{} by including the actually observed properties of spots and investigate what effect spots have on the \sind{} of the Sun and other, more active, stars. Therefore, in our current approach to synthesize the \sind{}, the total flux from the stellar disk at time $t$ in each of the four passbands is computed by combining the contributions from the quiet star with that from spots and faculae. The stellar disk is split into $l$ rings and the flux values calculated in these rings are summed up to obtain the total flux from the disk. Our approach can be summarised through the following equation for the total flux $N_{\rm m}(t)$:
\begin{eqnarray}
   N_{\rm m}(t) = \sum_l \bigg[\int_{\rm m} I^q(\lambda_{\rm m},\mu_l){\rm Tr(m)}\ d\lambda\bigg] \Delta\Omega_l  \nonumber && \\ +\sum_l \alpha^f_{l} (t)\bigg[\int_{\rm m}(I^f(\lambda_{\rm m},\mu_l) - I^q(\lambda_{\rm m},\mu_l)){\rm Tr(m)}\ d\lambda\bigg] \Delta\Omega_l\nonumber && \\ +\sum_l \alpha^s_{l}(t)\bigg[\int_{\rm m} (I^s(\lambda_{\rm m},\mu_l) - I^q(\lambda_{\rm m},\mu_l)){\rm Tr(m)}\ d\lambda\bigg]\Delta\Omega_l\ , \nonumber && \\
    \label{eq:flux}
\end{eqnarray}
with `m' representing H, K, R, and V passbands. $\alpha^f_{l}$ and $\alpha^s_{l}$ are the time-dependent disk area coverages of faculae and spots in a ring at a limb distance of $\mu_l$ which subtends a solid angle $\Delta\Omega_l$. $I$ are the intensities from a given feature that depend on wavelength, $\lambda_m$, and $\mu_l$. Tr(m) are the transmission profiles of the H, K, R, and V passbands.

The quiet star flux (first term in \eqn{eq:flux}) is determined using spectra synthesized from the standard 1D semi-empirical model C of \citet{Fontenlaetal1999} while facular and plage fluxes (second term in \eqn{eq:flux}) are obtained using spectra from model P of \citet{Fontenlaetal1999}. As explained in \citetalias{Sowmyaetal2021}, magnetic flux concentrations forming faculae on the photosphere undergo expansion in the chromosphere where they form plages. Due to this expansion, chromospheric plages cover larger areas on the Sun as compared to the photospheric faculae. In order to account for this expansion, we multiply faculae coverages $\alpha^f_{l}$ by a constant expansion factor while computing plage spectral fluxes. Essentially, the plage area coverages are taken to be 2.9 times the faculae area coverages. This value is slightly smaller than 3.2 used in \citetalias{Sowmyaetal2021} due to a number of minor adjustments in the model. Further details on the spectral synthesis and the use of a constant expansion factor are given in \citetalias{Sowmyaetal2021}.

We take the following approach for computing the spot contribution (last term in \eqn{eq:flux}) to the total disk-integrated flux. For the pseudo-continuum R and V fluxes, we utilize spectra representing the photospheres of umbral and penumbral regions within spots computed by \citet{famous_Yvonne_paper}. These spectra have been well tested against available solar irradiance measurements \citep{Sami2013,Ermolli2013}. The chromosphere above sunspots is either missing or not very well represented in the existing 1D models of umbra and penumbra \citep[see e.g.][]{Loukitchevaetal2017}. Therefore these models cannot be used to reliably determine the H and K fluxes of spots. Fortunately, in recent years, SST has observed sunspots in the Ca\,{\sc ii} spectral region with unprecedented spatial and good spectral resolution which makes it possible to study the chromospheric counterparts of spots in this spectral range. In the next section we detail these data and their analysis.

\subsection{Determining spot brightness in H and K passbands}
\label{ssec:sphk}
\begin{figure*}[ht!]
    \centering
    \includegraphics[scale=0.45]{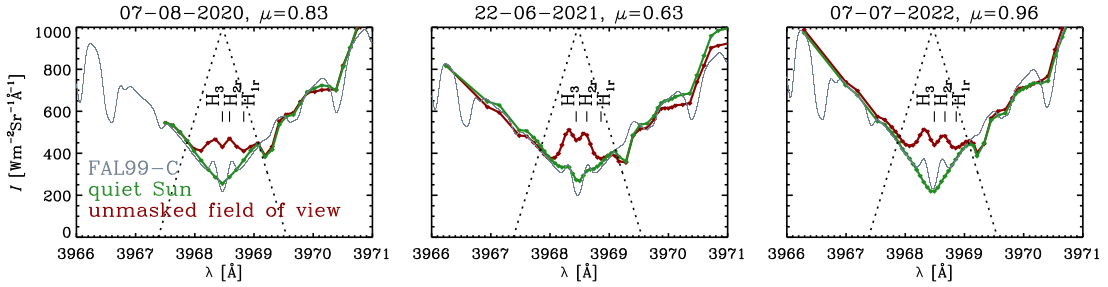}
    \caption{Average intensity profiles from the unmasked field of view in \fref{fig:intcombined} (maroon) and from the quiet Sun region marked by the green box in \fref{fig:intcombined} (green). Intensity profiles computed using the FAL99-C model of solar atmosphere (gray) are plotted for reference. The transmission profile of the H passband is shown by the dotted triangle. The H1r, H2r and H3 features are also marked.}
    \label{fig:profcombined} 
\end{figure*}

We analyse observations of sunspots in active regions (AR) AR12770, AR12833, and AR13046 obtained on 07-08-2020, 22-06-2021, and 07-07-2022, respectively, with CHROMIS at SST. CHROMIS is a dual Fabry-P\'erot filtergraph system which carries out fast wavelength sampling of spectral lines. The spectral bandwidth FWHM of CHROMIS is about 120\,m\AA{}. The imaging observations of the sunspot in AR12770 ($\mu={\rm cos}\ \theta=0.83$, with $\theta$ being the heliocentric angle) were taken with 29 spectral points in the H line with a step size of 120\,m\AA{}, including the central H3 absorption feature along with the blue(b)-red(r) H1 and H2 peaks. The spectral sampling was finer for the spots in AR12833 ($\mu=0.63$) and AR13046 ($\mu=0.96$), with 47 wavelength points separated by 60\,m\AA{} in most parts of the H line. All three spots were also scanned in the continuum around 4001\,\AA{}. The CHROMIS angular pixel scale is 0.038''. The observations cover a field of view of approximately 67''$\times$42'' and the duration of the spectral scan is between 10--15 seconds. The data were reduced following the CHROMISRED/SSTRED procedure outlined in \citet{Loefdahletal2021}.

Figure~\ref{fig:intcombined} shows observed images of the three spots at selected wavelengths. At chromospheric heights, the superpenumbrae of the spots become clearly visible (e.g. columns three and four). These superpenumbrae cover much larger areas than the spots at the photosphere. These superpenumbral regions are much brighter than the surrounding quiet regions (shown by green boxes in \fref{fig:intcombined}). A visual inspection of the Ca\,{\sc ii} H far wing intensity images showed very bright facular patches lying close to the sunspot in the observed field of view (e.g. bright features on the left in \fref{fig:intcombined}d). We mask these facular patches as shown in \fref{fig:intcombined}, and exclude them from the analysis. In each sunspot scan, we identify the quiet regions in the field of view (shown by green boxes in \fref{fig:intcombined}) and compute average intensity profiles, $\langle I\rangle_{\rm qs}$, from within these regions. These profiles are plotted in green in \fref{fig:profcombined}. We utilize spectra synthesized with the FAL99-C model to calibrate the observed quiet Sun profiles and determine the calibration factor. This factor is used to calibrate the entire image. We note that synthetic spectra computed with the FAL99-C model have been demonstrated to agree with FTS spectral atlas data \citep{Neckel1999,Doerretal2016}. We refer to \citetalias{Sowmyaetal2021} for details. 

Next we compute the average profile, $\langle I\rangle_{\rm ufov}$, from the unmasked field of view (ufov) shown in \fref{fig:intcombined}. The flux excess/deficit from the spot in the H passband is then calculated following
\begin{equation}
    P_{\rm ufov} =  \int_{\rm H} [\langle I\rangle_{\rm ufov} \cdot N_{\rm ufov}]\ d\Omega\ {\rm Tr(H)}\ d\lambda\ ,
    \label{eq:spot}
\end{equation}
\begin{equation}
    P_{\rm qs} =  \int_{\rm H} [\langle I\rangle_{\rm qs} \cdot N_{\rm ufov}]\ d\Omega\ {\rm Tr(H)}\ d\lambda\ ,
    \label{eq:qs}
\end{equation}
\begin{equation}
    \Delta{\rm spot} =  P_{\rm ufov} - P_{\rm qs}.
    \label{eq:deltaspot}
\end{equation}
Here $N_{\rm ufov}$ is the total number of pixels within the unmasked field of view and $d\Omega$ is the solid angle subtended by a pixel. $P_{\rm ufov}$ is the flux from the unmasked field of view which contains both spot and quiet regions. $P_{\rm qs}$ is the flux which would be measured if the entire unmasked field of view were covered by the quiet Sun. When the difference between these fluxes is calculated in \eqn{eq:deltaspot} the contribution of pixels corresponding to the quiet Sun gets cancelled out. Thus, what remains is the flux excess/deficit due to the spot, including the canopy.

The flux difference due to spots in W\,m$^{-2}$ units calculated using \eqn{eq:deltaspot} is given in the fourth column of \tab{tab:sp}. The positive flux difference values indicate that the chromospheric counterparts of spots are brighter than the quiet surroundings, which is already clear from \fref{fig:profcombined} where the mean profiles from the unmasked field of view are much brighter in the line core compared to the mean quiet Sun profiles (compare maroon and green profiles). These flux values depend on the size of the sunspot. In order to remove this dependency, we divide the flux values by the projected sunspot area on the photosphere, i.e., the area enclosed within the black contours in \fref{fig:intcombined}. The fifth column of \tab{tab:sp} provides these area normalised fluxes. For comparison, the facular fluxes per unit area from \citetalias{Sowmyaetal2021} are also given in \tab{tab:sp}. While somewhat lower, the sunspot brightness in H passband is still of comparable magnitude to the facular brightness.

\begin{table*}[ht!]
\centering
\caption{Spot brightness in H passband in comparison to the surrounding quiet Sun regions, calculated directly from SST data. The model values from \citetalias{Sowmyaetal2021} for faculae are shown in the last column for reference.}
\begin{tabular}{c|c|c|c|c||c}
\midrule[1.5pt]
\multirow{2}{*}{DD-MM-YYYY} & \multirow{2}{*}{Active region} & \multirow{2}{*}{$\mu$} & \multirow{2}{*}{\mbox{$\Delta$spot} (W\,m$^{-2}$)} & $\Delta${\rm spot} (W\,m$^{-2}$) & $\Delta${\rm faculae} (W\,m$^{-2}$)\\
& & & & $\overline{{\rm spot\ area}\ ({\rm km}^2)}$ & $\overline{{\rm faculae\ area}\ ({\rm km}^2)}$\\
\midrule[1.5pt]
07-08-2020 & 12770 & 0.83 & 3.93e-06 & 1.94e-14 & 6.53e-14\\
22-06-2021 & 12833 & 0.63 & 4.56e-06 & 1.32e-14 & 6.38e-14 \\
07-07-2022 & 13046 & 0.96 & 5.80e-06 & 2.04e-14 & 6.62e-14 \\
\bottomrule[1.5pt]
\end{tabular}
\label{tab:sp}
\end{table*}

\begin{figure}[ht!]
    \centering
    \includegraphics[scale=0.4]{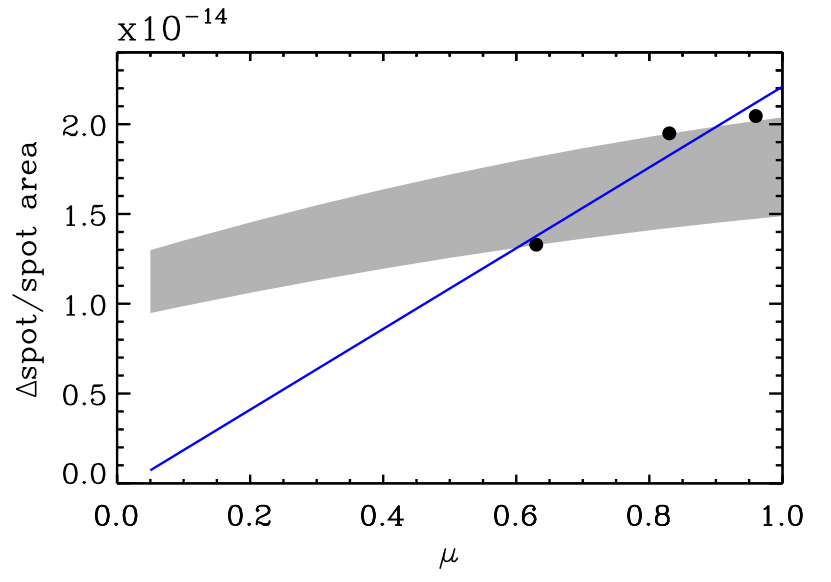}
    \caption{Center to limb variation of the flux excess due to spots. Black circles are the SST observations (values from fifth column of \tab{tab:sp}). Grey shaded region shows different scalings of the quiet Sun CLV to match observations. Blue line is the linear fit to the observed data points. See text for details.}
    \label{fig:sphclv} 
\end{figure}

CHROMIS observations analyzed in this study cover $\mu$-values from 0.96 down to 0.63 which corresponds to more than 50\% of the solar disk. Nevertheless, to compute the spot contribution to the total disk-integrated flux in the H passband, we require spot contrasts outside the $\mu$-range covered by these observations.  To obtain them, we make several experiments assuming various dependence of spot contrast on $\mu$. First, we make use of the centre to limb variation (CLV) of quiet Sun flux in the H passband calculated using the FAL99-C spectra. Earlier studies have shown that the \ca{} emission from the quiet Sun and faculae have the same CLV \citep[e.g.][]{Skumanichetal1984,Ermollietal2007,Ermollietal2010}. Following this, we assume that the \ca{} emission from spots also has the same CLV as the quiet Sun. Then we use $\mu=0.84$ or $\mu=0.63$ observations to define the range of CLV dependences of spot contrast considered for \sind{} calculations (see grey shaded area in \fref{fig:sphclv}). We also perform a linear regression fit to the observed data points which results in a much steeper CLV as shown by the blue line in \fref{fig:sphclv}. Despite such a wide range of considered CLV dependences the resulting \sind{} values are very similar (see \sref{ssec:as}) so that the exact behaviour of the spot contrast outside the observed $\mu$-range has very little impact on the results of this study.

Further, we expect the flux difference due to spots in the K passband to be similar to that in the H passband. This is because the \ca{} lines have similar excitation energies and Einstein coefficients so that they form at nearly the same heights in the solar atmosphere and also respond similarly to the changes in the atmospheric structure associated with magnetic activity \citep[e.g.][]{Uitenbroek1990}. Therefore we use the CLV curves shown in \fref{fig:sphclv} also for the K passband. Furthermore, small sunspots with relatively simple morpholigies similar to the ones shown in \fref{fig:intcombined} are found to be a lot more common than big sunspots \citep{Bogdanetal1988,BaumannSolanki2005} that can have a more complex morphology. Hence, we expect that the morphology of sunspots shown in \fref{fig:intcombined} is representative of a good fraction of sunspots. Consequently, we use the same CLV for spots of all shapes and sizes.

\section{Influence of spots on the \texorpdfstring{$S$}{}-index}
\label{sec:res}
We now compute the \sind{} using the method described in \sref{ssec:model}. First, we revisit the assumption made in \citetalias{Sowmyaetal2021} that spot contribution to the solar \sind{} is negligible. Then we consider active stars with larger spot coverages than the Sun and evaluate, for the first time, how the \sind{} is affected by spots.

\begin{figure*}[ht!]
    \centering
    \includegraphics[scale=0.55]{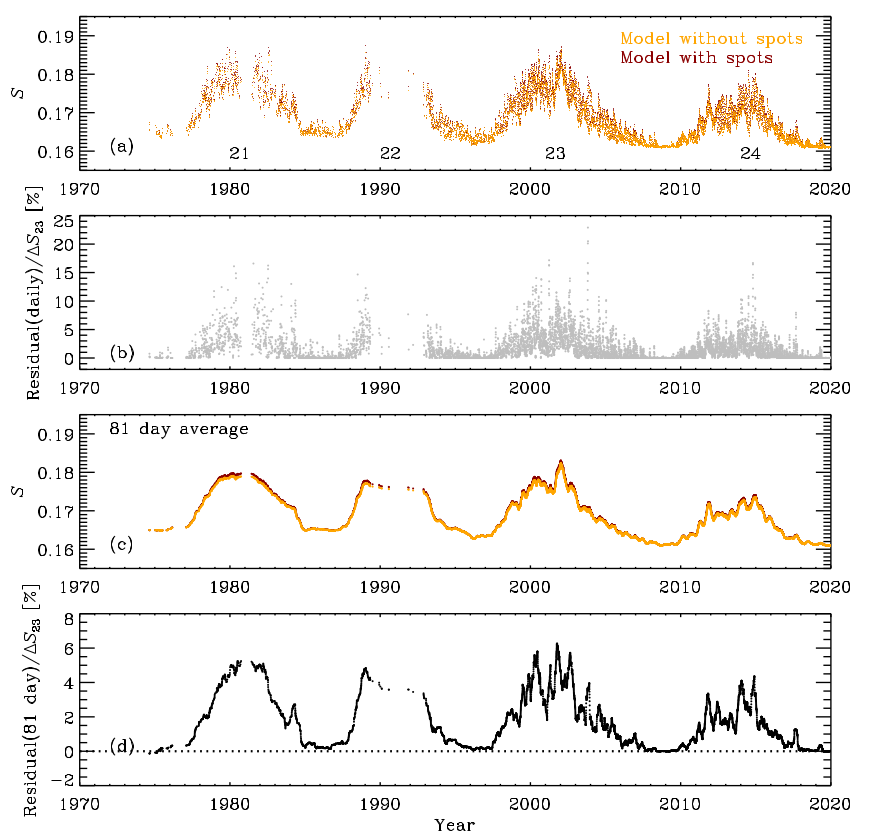}
    \caption{Panel a: comparison between the solar \sind{} values computed including (maroon) and excluding (orange) the contribution from spots. Panel b: difference between the daily \sind{} values shown in panel (a) normalised to the amplitude of the solar cycle 23 in \sind{}, which is 0.0149. Panel c: 81 day smoothed values of the \sind{} time series plotted in panel (a). Panel d: difference between the 81 day averaged values shown in panel (c) normalised to the amplitude of the solar cycle 23 in \sind{}.}
    \label{fig:sindex} 
\end{figure*}

\subsection{The Sun}
\label{ssec:Sun}
\fref{fig:sindex} shows how sunspots influence the observed solar \ca{} emission. A comparison of the daily \sind{} timeseries for four solar activity cycles computed by including the contribution from spots and neglecting it  is shown in Figures~\ref{fig:sindex}a-b, while Figures~\ref{fig:sindex}c-d show the 81-day smoothed values. We only present the calculations done using the CLV corresponding to the upper envelope of the grey region shown in \fref{fig:sphclv}. This is because the solar \sind{} values computed for the range of CLVs shown in \fref{fig:sphclv} do not differ much. The area coverages of faculae and spots needed for the \sind{} computations are taken from \citet{Yeoetal2014}, see \citetalias{Sowmyaetal2021} for details.

Sunspots are dark in the photosphere compared to the quiet Sun so that they decrease solar flux in the R and V passbands. At the same time, the chromospheric counterparts of spots are brighter than the surrounding quiet regions so that spots enhance the total flux in the H and K passbands. Both of these effects lead to an increase of the \sind{} when sunspots are included in the calculations. The contribution of spots to \sind{} is, unsurprisingly, zero at minima of solar activity and increases towards activity maxima. Figure~\ref{fig:sindex}b shows that transits of large spot groups over the visible solar disk cause short-term increases of the \sind{} by up to 25\,\%\ of the amplitude of the solar cycle~23 in \sind{}.  At the same time Figure~\ref{fig:sindex}d shows that the effect of spots on 81-day smoothed values is significantly less and is below 7\,\%\ of the amplitude of the solar cycle 23 in \sind{}.

All in all, since disk area coverages of solar faculae are much larger than those of spots and since faculae are a bit brighter than spots in the H and K passbands (see \tab{tab:sp}), the variations of the solar \sind{} are almost fully brought about by faculae with exceptions of periods when large sunspot groups transit the visible solar disk.

\subsection{Active stars}
\label{ssec:as}
\begin{figure*}
    \centering
    \includegraphics[scale=0.45]{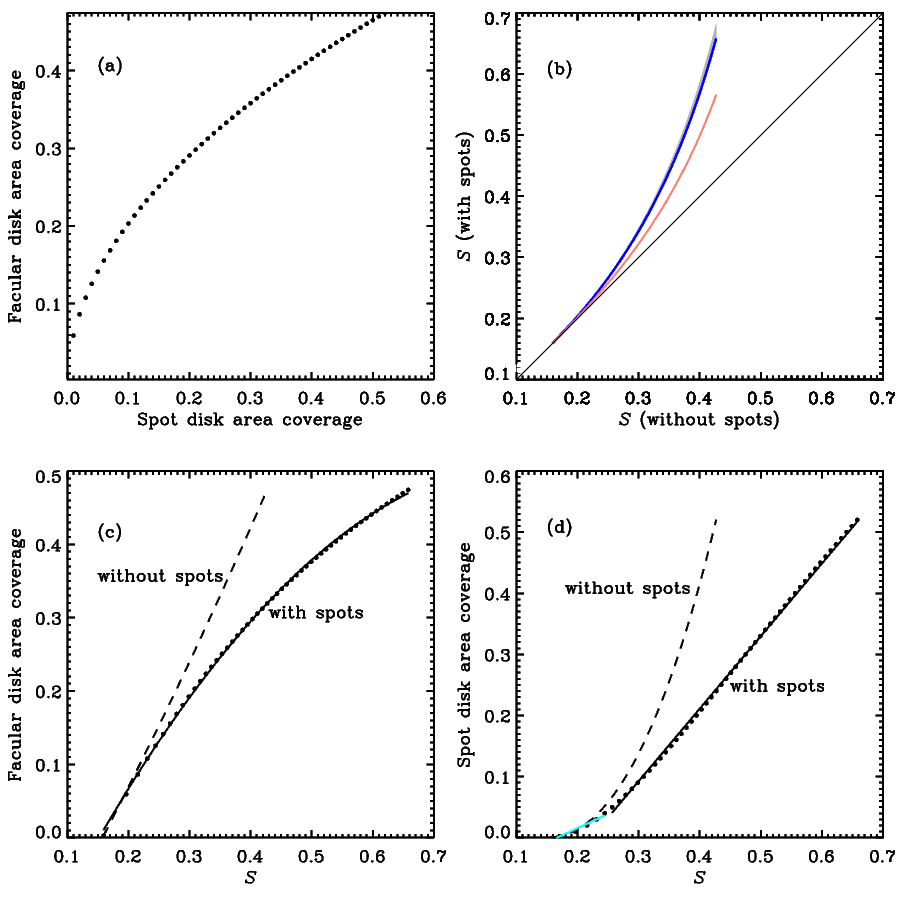}
    \caption{Panel a: photospheric disk area coverages of faculae as a function of the corresponding spot area coverages, see text for details. Panel b: \sind{} with spots as a function of \sind{} without spots. The blue curve and the grey shaded area are obtained using the spot contrast CLVs shown in \fref{fig:sphclv}. The salmon curve represents the case where we assume that spots are as bright as the quiet star in H and K so that they are invisible in the chromosphere but are still visible as dark features in the photosphere. The black line shows the case when spots are neglected completely. Panel c: relationship between facular area coverages and \sind{} with spots (filled circles) along with the quadratic fit to this relationship (solid line). The dashed line shows the dependence of facular disk area coverage on \sind{} without spots. Panel d: relationship between spot area coverages and \sind{} with spots (filled circles) along with the linear fits (cyan and black solid lines) to their relationship. The dashed line shows the dependence of spot disk area coverages on \sind{} without spots.}
    \label{fig:activesindex} 
\end{figure*}

A number of studies have shown that solar spot coverages increase with solar activity faster than those of faculae  \citep[e.g.][]{Foukal1993,Chapmanetal1997,Foukal1998,Yeoetal2021,Theoetal2022}. \cite{Sashaetal2014} and \cite{Nemecetal2022} showed that the solar relation between facular and spot coverages can be extrapolated towards stars more active than the Sun. This implies that the ratio of spot-to-facular coverages is higher for active stars and, thus, the relative contribution of spots to their \sind{} values is expected to be higher than that reported for the Sun in \sref{ssec:Sun}. 

To find the spot contribution to \sind{} of active stars we use the relationship between the spot and facular area coverages established based on the solar data \citep{Sashaetal2014, Ninathesis2021}:
\begin{equation}
    \alpha^f = a \sqrt{\alpha^s-b} + c\ ,
    \label{eq:fillfactor}
\end{equation}
where $a,b,c$ are constants. Their values are 0.67, $-0.0003$, and $-0.009$, respectively. Here we assume that 
this relationship holds also for more active stars \citep[see detailed discussion in][]{Nemecetal2022} and consider a series of spot area coverage values from 0 to 50\,\%. Then we use \eqn{eq:fillfactor} to get the corresponding facular area coverages. The facular area coverages obtained in this way are shown as a function of spot area coverages in \fref{fig:activesindex}a. For simplicity, we assume that spots and faculae are uniformly distributed on the stellar disk and use the model described in \sref{sec:methods} to calculate \sind{} values.
 
Figure~\ref{fig:activesindex}b shows the \sind{} values calculated by including both faculae and spots plotted as a function of \sind{} values computed by completely neglecting spots \citepalias[or rather assuming that they are equally dark in \ca{} line cores and continuum as was done in][]{Sowmyaetal2021}. The blue curve and the grey shaded area represent the cases when the spot contribution is taken into account both in the line cores (H and K) and in the continuum (R and V). For the line cores, we use the entire range of CLVs shown in \fref{fig:sphclv} while for the continuum, we use the CLV computed using synthetic spectra from \citet{famous_Yvonne_paper}. We see from \fref{fig:activesindex}b that different CLVs of the line core contrasts lead to very small changes in the \sind{} values, which is reassuring. It is also clear from this plot that the contribution from spots to \sind{} becomes progressively more significant with increasing spot coverages. Spots lead to a steep increase of the \sind{} by decreasing the radiative flux in the continuum and increasing it in the \ca{} line cores.

Further, we do an experiment to understand which of the two effects from spots, namely the reduction in the continuum flux and enhancement of the line core flux, is the dominant one. In this experiment the chromospheric counterparts of spots are assumed to be as bright as the quiet chromosphere so that they do not affect the line core flux. However, the effect of spots on the continuum flux is still accounted for. The resulting \sind{} values from this experiment are shown by the salmon curve in \fref{fig:activesindex}b. A comparison of the salmon curve with the other cases shown in \fref{fig:activesindex}b suggests that the effect arising from the decrease of continuum fluxes is slightly larger than that from the increase of the flux in the line core. All in all, the contribution of spots to the \sind{} cannot be neglected for active stars, e.g. the error in the \sind{} resulting from neglecting spots reaches 16\,\%\ for $S=0.3$ and 30\,\%\ for $S=0.35$.

In Figures~\ref{fig:activesindex}c-d, we show the relationship between the \sind{} computed with spots (representing the upper envelope of the grey shaded area in \fref{fig:activesindex}b) and disk area coverages of spots and faculae. Facular disk area coverages show a quadratic dependence on the \sind{} while the spot disk area coverages increase linearly with the \sind{}. We find the following dependence for the faculae disk area coverages:
\begin{equation}
    \alpha^f(S) = -0.246 + 1.773\ S - 1.032\ S^2\ ,
\end{equation}
and for the spot disk area coverages:
\begin{align}
\alpha^s(S) = \begin{cases}
-0.078 + 0.466 S & \text{if $S < 0.25$}\ ; \\
-0.269 + 1.209 S & \text{if $S \ge 0.25$}\ ,
\end{cases}
\end{align} 
where $S$ here is the \sind{} computed by including spots.

Figures~\ref{fig:activesindex}c-d point to a linear relationship for faculae disk area coverages and quadratic relationship for spot disk area coverages as a function of \sind{} when spot contribution to the \sind{} is neglected. The same functional relationships (see their Eqs.~1--2) have been found by  \citet{Sashaetal2014}  using solar data, i.e. in the regime when \sind{} variability is mainly driven by faculae. Our result shows that the spot contribution to the \sind{} for active stars changes the functional form of the dependence of the facular and spot disk area coverages on the \sind{}.

\section{Summary and Discussion}
\label{sec:conclu}
While empirical and physics-based models attribute faculae/plage to be the main contributors to the \sind{} thereby neglecting spots, the role played by spots was until now poorly understood. In this paper, for the first time, we explored the effect of spots on stellar \ca{} emission. For this purpose, we used high quality observations of three sunspots from SST. The analysis of these sunspots revealed that in the \ca{} line cores, sunspots are brighter than the quiet Sun regions, thus leading to a positive contrast value in the H and K passbands. The sunspots are nearly as bright as faculae in these spectral passbands. Spots reduce the total flux in the R and V passbands because they are dark in the photosphere while at the same time they increase the total H and K flux. Together, the two effects lead to an enhancement in the \sind{}.

For an old, less-active star like the Sun, disk area coverages of spots are significantly smaller than those of faculae. Because of this, the increase in the \sind{} caused by sunspots is only a small fraction of the observed solar \sind{} and its variability. Therefore, we confirm that neglecting the spot contribution is a reasonable choice when modeling the \sind{} of inactive stars like the Sun. However, we find that spots have a great impact on the \sind{} for young and active stars. Thus, spot contribution cannot be neglected when interpreting measurements of the \sind{} and its variability in active stars. We also find, under the assumptions of the modelling approach we used, that for active stars the facular disk area coverages show a quadratic dependence on the \sind{} while the dependence for spot disk coverage is linear.

An important limitation of our modeling approach is that it does not allow accounting for the interaction between chromospheric counterparts of different active regions. Along the same line, chromospheric superpenumbrae of different spots within an active region can also interact with each other if the spots get packed too tightly. These effects are outside of the scope of this study and warrant further investigation.

Although metrics based on stellar photometric brightness have recently emerged as magnetic activity indicators \citep[e.g.][]{Basrietal2013}, the \sind{} still serves as an important proxy for stellar activity. A plethora of \sind{} data has recently been collected by surveys targeting exoplanets. For example, the Keck and Lick observatories have recorded $S$-indices of over 2000 FGKM dwarfs as part of the California Planet Search program \citep{Wrightetal2004,IsaacsonandFischer2010}. The High Accuracy Radial velocity Planet Searcher \citep[HARPS;][]{Mayoretal2003} provides $S$-indices of over 1600 FGK stars \citep[e.g.][]{Gomesetal2021}. Observations of the chromospheric emission from stars, including extended monitoring of those in the Mount Wilson HK sample, are expected from the EXtreme PREcision Spectrometer \citep[EXPRES;][]{Jurgensonetal2016} at Lowell observatory. Furthermore, the spectroscopic survey by the Large Sky Area Multi-Object Fibre Spectroscopic Telescope \citep[LAMOST;][]{Cuietal2012,Zhaoetal2012} has allowed measuring chromospheric activity of hundreds of thousands of stars \citep{LAMOSTachievements, lamostsindex2022}. All these measurements show a considerable spread in the \sind{} distribution of stars belonging to a given spectral type. An understanding of the main drivers of the chromospheric activity of these stars will greatly benefit the interpretation of their \sind{} measurements.

Here we considered the effect of spots on \ca{} emission from stars with solar fundamental parameters. In the forthcoming publication we plan to study the effect of spots on activity cycles of stars using the surface flux transport simulations \citep{Isiketal2018,Nemecetal2023}.

\begin{acknowledgements}
We thank the anonymous referee for their helpful comments. The Swedish 1-m Solar Telescope is operated on the island of La Palma by the Institute for Solar Physics of Stockholm University in the Spanish Observatorio del Roque de los Muchachos of the Instituto de Astrof{\'\i}sica de Canarias. The Institute for Solar Physics is supported by a grant for research infrastructures of national importance from the Swedish Research Council (registration number 2021-00169). LRvdV is supported by the Research Council of Norway, project number 325491, and through its Centres of Excellence scheme, project number 262622. This study has made use of SAO/NASA Astrophysics Data System's bibliographic services.
\end{acknowledgements}

\bibliographystyle{aasjournal}
\bibliography{caiihspot}

\end{document}